\newif\ifsubmission
\newif\ifanonymous
\newif\ifcomments
\newif\ifresponse

\ifdefined\issubmission
\submissiontrue
\fi

\ifdefined\isanonymous
\anonymoustrue
\fi

\ifdefined\iscomments
\commentstrue
\fi

\ifdefined\isresponse
\responsetrue
\fi

\ifcomments
\newcommand\sandeep[1]{\textcolor{blue}{Sandeep: #1}}
\newcommand\asokan[1]{\textcolor{red}{Asokan: #1}}
\newcommand\eledra[1]{\textcolor{green}{Long: #1}}
\newcommand\andrew[1]{\textcolor{orange}{Andrew: #1}}
\else

\newcommand\sandeep[1]{}
\newcommand\asokan[1]{}
\newcommand\eledra[1]{}
\newcommand\andrew[1]{}
\fi

\ifresponse
\newcommand\revised[1]{{\leavevmode\color{blue}{#1}}}
\else
\newcommand\revised[1]{{#1}}
\fi

\newcommand{\OmniShare}{{OmniShare}\xspace}
\newcommand{\NewDev}{\textsfsl{A}\xspace}
\newcommand{\AuthDev}{\textsfsl{B}\xspace}

\newcommand{\filekey}{\textit{filekey}\xspace}

\newcommand{\folder}{directory\xspace}
\newcommand{\folders}{directories\xspace}
\newcommand{\Folder}{Directory\xspace}

\DeclareTextFontCommand{\textsfsl}{\sffamily\slshape}

\ifsubmission
\documentclass[letterpaper, journal]{IEEEtranSub}
\usepackage{geometry}
\else
\documentclass[letterpaper, journal]{IEEEtran}
\fi

\usepackage[hidelinks]{hyperref}

\usepackage[T1]{fontenc}
\usepackage{bm}
\DeclareMathAlphabet{\mathsfit}{\encodingdefault}{\sfdefault}{m}{sl}
\SetMathAlphabet{\mathsfit}{bold}{\encodingdefault}{\sfdefault}{bx}{sl}

\newcommand{\tens}[1]{{\mathsfit{#1}}}
\usepackage{cite}

\ifCLASSINFOpdf
  \usepackage[pdftex]{graphicx}
\else
\fi

\usepackage{listings}
\usepackage{xcolor}

\usepackage{xspace}

\colorlet{punct}{red!60!black}
\definecolor{background}{HTML}{EEEEEE}
\definecolor{delim}{RGB}{20,105,176}
\colorlet{numb}{magenta!60!black}

\lstdefinelanguage{json}{
    basicstyle=\normalfont\ttfamily,
    numbers=left,
    numberstyle=\scriptsize,
    stepnumber=1,
    numbersep=8pt,
    showstringspaces=false,
    breaklines=true,
    frame=lines,
    backgroundcolor=\color{background},
    literate=
     *{0}{{{\color{numb}0}}}{1}
      {1}{{{\color{numb}1}}}{1}
      {2}{{{\color{numb}2}}}{1}
      {3}{{{\color{numb}3}}}{1}
      {4}{{{\color{numb}4}}}{1}
      {5}{{{\color{numb}5}}}{1}
      {6}{{{\color{numb}6}}}{1}
      {7}{{{\color{numb}7}}}{1}
      {8}{{{\color{numb}8}}}{1}
      {9}{{{\color{numb}9}}}{1}
      {:}{{{\color{punct}{:}}}}{1}
      {,}{{{\color{punct}{,}}}}{1}
      {\{}{{{\color{delim}{\{}}}}{1}
      {\}}{{{\color{delim}{\}}}}}{1}
      {[}{{{\color{delim}{[}}}}{1}
      {]}{{{\color{delim}{]}}}}{1},
}

\usepackage{url}

\usepackage{longtable}

\begin{document}
%
\title{OmniShare: Securely Accessing Encrypted Cloud Storage from
Multiple Authorized Devices}
%
%
%

\ifanonymous

\else{
\author{\IEEEauthorblockN{Andrew Paverd\IEEEauthorrefmark{1},
                          Sandeep Tamrakar\IEEEauthorrefmark{1},
                          Hoang Long Nguyen\IEEEauthorrefmark{2},
                          Praveen Kumar Pendyala\IEEEauthorrefmark{3},
                          Thien Duc Nguyen\IEEEauthorrefmark{3},
                          Elizabeth Stobert\IEEEauthorrefmark{4},
                          Tommi Gr\"{o}ndahl\IEEEauthorrefmark{1},
                          N. Asokan\IEEEauthorrefmark{1},
                          and
                          Ahmad-Reza~Sadeghi\IEEEauthorrefmark{3}}
  
~

        \IEEEauthorblockA{\small\IEEEauthorrefmark{1}Aalto University, \texttt{andrew.paverd@ieee.org}, \{\texttt{sandeep.tamrakar, tommi.grondahl}\}\texttt{@aalto.fi}, \texttt{asokan@acm.org}}\\
        \IEEEauthorblockA{\small\IEEEauthorrefmark{2}LORIA, Universit\'{e} de Lorraine/INRIA/CNRS, \texttt{hoang-long.nguyen@loria.fr}}\\
        \IEEEauthorblockA{\small\IEEEauthorrefmark{3}Technische Universit\"{a}t  Darmstadt, \{\texttt{praveen.pendyala, ducthien.nguyen, ahmad.sadeghi}\}\texttt{@trust.tu-darmstadt.de}}\\
        \IEEEauthorblockA{\small\IEEEauthorrefmark{4}ETH Z\"{u}rich, \texttt{estobert@inf.ethz.ch}}
%
	}

%
%

\fi

%



\ifresponse
\input{contents/IC_changelog.tex}
\fi

\maketitle

\begin{abstract}
Cloud storage services like Dropbox and Google Drive are widely used by individuals and businesses.
Two attractive features of these services are 1)~the automatic synchronization of files between multiple client devices and 2)~the possibility to share files with other users.
However, privacy of cloud data is a growing concern for both individuals and businesses.
Encrypting data on the client-side before uploading it is an effective privacy safeguard, but it requires all client devices to have the decryption key. 
Current solutions derive these keys solely from user-chosen passwords, which have low entropy and are easily guessed.

We present \OmniShare, the first scheme to allow client-side encryption with high-entropy keys whilst providing an intuitive key distribution mechanism to enable access from multiple client devices.
Instead of passwords, we use low bandwidth uni-directional out-of-band (OOB) channels, such as QR codes, to authenticate new devices.
To complement these OOB channels, the cloud storage itself is used as a communication channel between devices in our protocols.
We rely on a directory-based key hierarchy with individual file keys to limit the consequences of key compromise and allow efficient sharing of files without requiring re-encryption.
\OmniShare is open source software and currently available for Android and Windows with other platforms in development.
We describe the design and implementation of \OmniShare, and explain how we evaluated its security using formal methods, its performance via real-world benchmarks, and its usability through a cognitive walkthrough.

\end{abstract}

\ifsubmission
\begin{IEEEkeywords}
Access controls, Storage/repositories, Data sharing
\end{IEEEkeywords}
\fi

%
\IEEEpeerreviewmaketitle

\section{Introduction}
\label{sec:intro}

Cloud storage services, such as Dropbox and Google Drive, are increasingly being used by individuals and businesses.
\revised{The results of the 2014 European \emph{Survey on ICT usage in households and by individuals} show that one in every five people used cloud storage services in 2014~\cite{ICT2014}.
The two foremost reasons for using cloud services, as given by current users, were:
\begin{itemize}
\item The possibility to use files from several devices or locations (cited by 59\% of users)
\item The ability to easily share files with other users (cited by 59\% of users)
\end{itemize}
However, concerns about data privacy are limiting the uptake of cloud storage services. 
Of the respondents who had used the internet and were aware of cloud services but did not use them, security or privacy concerns were given as the main reason for not using these services (cited by 44\% of respondents in this category)~\cite{ICT2014}.}
Although all major cloud storage providers use secure communication channels and routinely encrypt data before storing it, the original data is still available to the service providers themselves. 
Anyone with access to the service provider's infrastructure, legitimately or otherwise, can read and modify
this data, often without detection~\cite{soghoian:cloud}.
For individuals, this could lead to loss of privacy and identity theft, whilst for businesses, this could have legal consequences.

\revised{Encrypting data on the client-side before uploading it to the cloud is an effective way to mitigate this risk.
However, to retain the first main benefit cited above, this data must be accessible from \emph{all} the user's devices.
For example, assume Alice encrypts a file on her PC and uploads it to Dropbox.
If she wants to access this file from her smartphone, Alice's smartphone must have (or be able to obtain) the relevant decryption key.
Naturally, these keys cannot be managed by the cloud service provider, thus resulting in a key distribution problem.
Current encrypted storage services, such as SpiderOak~\cite{SpiderOak} and Tresorit~\cite{Tresorit}, address this problem by deriving keys from the user's password using a deterministic password-based key derivation function (PBKDF).
Alice's smartphone can derive the relevant keys from her password.
However, it is well-known that human-chosen passwords usually have very low entropy and are easily guessed.
Through analysis of a 70~million password corpus, Bonneau estimated that human-chosen passwords provide only about 20~bits of security against an optimal \emph{offline} dictionary attack~\cite{Bonneau2012}. 
This is significantly less security than the cryptographic keys used in any modern encryption system.
An adversary who can obtain Alice's encrypted files, including the cloud storage provider itself, is theoretically capable of performing this type of attack.
The analysis also showed that, even for more security-sensitive tasks, users do not choose significantly stronger passwords~\cite{Bonneau2012}.}
To avoid deriving keys from passwords, services such as Viivo~\cite{Viivo}, BoxCryptor~\cite{BoxCryptor}, and Sookasa~\cite{Sookasa} use additional servers to manage and distribute keys, but this adds cost and introduces new vulnerabilities.

We present \OmniShare, the first scheme to allow client-side encryption with high-entropy keys whilst providing an intuitive key distribution mechanism to enable access from multiple client devices.
Instead of deriving keys from potentially weak passwords, \OmniShare encrypts files using high-entropy keys that are generated on the client devices (possibly within on-board secure hardware).
To enable access to these encrypted files from multiple authorized devices, the user can create an \OmniShare \emph{domain} to represent a group of devices.
All devices in the user's domain have access to the relevant encryption and decryption keys.
Instead of using additional trusted servers to distribute keys, \OmniShare uses a novel combination of an out-of-band (OOB) channel and the cloud storage service itself to distribute these keys.
To simplify this process for the user, \OmniShare automatically selects a suitable OOB channel between the new device and a previously authorized device based on their hardware capabilities.
Although OOB key distribution is not in itself a new idea (e.g. \cite{safeslinger}), to the best of our knowledge this approach has not yet been applied to the challenge of secure yet usable cloud storage.
This is not a straight-forward application of OOB key distribution.
For example, compared to generic OOB key distribution, this scenario gives rise to certain new requirements, such as the need to distribute many keys (e.g. one per file) and the need to share keys between multiple devices belonging to the same user.
Furthermore, it also provides certain unique capabilities, such as the ability for devices to use the cloud storage service itself as a communication channel - a feature utilized in \OmniShare.
Therefore, the main contribution of our work is the application and evaluation of OOB key distribution to the specific use case of encrypted cloud storage.

We have analysed the security of \OmniShare's protocols using formal methods (Section~\ref{sub:evaluation-protocol}).
\OmniShare also allows users to share encrypted files with other users.
By using a directory-based key hierarchy and encrypting each file with a unique key, \OmniShare can perform this sharing without requiring re-encryption of files.
We evaluated the performance of our implementation through real-world benchmarks (Section~\ref{sub:evaluation-performance}).
Usability is a primary consideration and thus the design of \OmniShare minimizes the amount of user interaction required during the authorization protocols and provides a consistent user experience across platforms and authorization mechanisms.
We evaluated the usability of \OmniShare by means of a \emph{cognitive walkthrough} (Section~\ref{sub:evaluation-usability}).
The aim of this evaluation was not to compare \OmniShare to other (less secure) systems, but rather to identify any obstacles that could inhibit a new user from learning to use this system.
As a generic means of securely yet intuitively defining domains of devices, \OmniShare can also be used for authentication and access control in other applications beyond secure cloud storage e.g., encrypted messaging services. 
\OmniShare is open source software available under the Apache 2.0 license.
It is currently available for Windows and Android with an iOS version in progress\footnote{\url{https://ssg.aalto.fi/projects/omnishare/}}.



\section{Requirements} 
\label{sec:requirements} 

We first define our adversary model and use it to identify the relevant security, functional and usability requirements. 

\noindent \textbf{Adversary Model}

We assume that the adversary can access, add and modify files in the user's cloud storage (i.e. the cloud storage provider itself may be the adversary). 
We do not attempt to protect against denial of service, e.g. deleting files in cloud storage. 
We assume that the adversary may collude with other users, including those with whom the primary user chooses to share files.
However, the adversary cannot observe or interfere with the \emph{local} interactions between the devices and users, which is a reasonable assumption given the network-oriented nature of the adversary.

\noindent \textbf{Functional Requirements}
\begin{enumerate}
	 \renewcommand{\labelenumi}{F\arabic{enumi}. }
	 \item All the user's authorized devices (i.e. devices in the user's \OmniShare \emph{domain}) must be able to access the user's encrypted files and directories.
	 \item Once a device has been added to an \OmniShare domain, it must be able to access the encrypted files without requiring any further interaction with other devices.
	 \item Users must be able to selectively share of individual files with other users.
	 \item \OmniShare should not be limited to a specific cloud storage provider.
\end{enumerate}

\noindent \textbf{Security Requirements}

\begin{enumerate}
	 \renewcommand{\labelenumi}{S\arabic{enumi}. }
	 \item Files must be encrypted using high-entropy keys generated on client devices before being uploaded to the cloud.
	 \item The decryption keys must only be accessible to devices within the \OmniShare domain.
\end{enumerate}

\noindent \textbf{Usability Requirements}
\begin{enumerate}
	 \renewcommand{\labelenumi}{U\arabic{enumi}. }
	 \item User actions during device authorization should be minimal, intuitive, and consistent across platforms and mechanisms.
\end{enumerate}

\ifsubmission
\section{Architecture and Implementation}
\label{sec:architecture}

We have implemented \OmniShare on Windows and Android, and support Dropbox as the cloud storage service.
Support for other platforms and cloud storage providers can be added without modifications to the architecture (Requirement F4). 
We provide specific details of our implementation alongside the \OmniShare architecture in this section.

\else
\section{Architecture}
\label{sec:architecture}

\fi

\OmniShare is designed as an application that runs on client devices. 
Users link their cloud storage to the application when it first runs on a new device.
Each device is assumed to have a \emph{device keypair} and a device-specific \emph{authentication key}, both of which could be protected by on-board secure hardware.
When a user initializes an \OmniShare domain, the application adds the initial device as the first authorized device. 
Creating a domain involves creating an \OmniShare \emph{directory}, a \emph{domain descriptor} file in this directory, and a \emph{root key} for the domain.
All files managed by \OmniShare are stored under the \OmniShare directory so that the user can also store non-encrypted files on the same cloud storage outside this directory.
The domain descriptor file records the following metadata for each authorized device:

\begin{itemize}
\item device name and unique identifier
\item available hardware capabilities
\item the device's public key
\item the domain root key (encrypted with the device public key) and the associated message authentication code (MAC) calculated using the authentication key.
\end{itemize}
The hardware capabilities are a list of the available peripherals that can be used for user input/output such as the device's camera (input), display (output), Near Field Communication (NFC) (input/output), and keyboard (input).
The domain descriptor file is not security sensitive and may be accessed/modified by the adversary.

\subsection{File Encryption and Key Hierarchy}
\label{sub:fileEncnCrypto}

When a file is first created, \OmniShare generates a new file key to encrypt the file.
\ifsubmission
In our implementation, files are encrypted using the Advanced Encryption Standard in Galois Counter Mode (AES-GCM) with 128-bit keys, but any semantically secure encryption scheme could be used.
\else
\fi
Encrypting each file separately allows \OmniShare to selectively share individual files with other users (Requirement F3).
\OmniShare also establishes and maintains a \emph{key hierarchy} with levels corresponding to the subdirectories of the \OmniShare directory.
Keys at each level are encrypted by the key of the level above, or by the root key for top level directories.
\OmniShare encrypts the root key separately with the public key of each authorized device using a \emph{lock-box} data structure~\cite{lockbox} (Requirement S2).

\begin{figure}[t]
   \centering
   \includegraphics[trim=1cm 1cm 2cm 1cm, clip=true, width=1\linewidth]{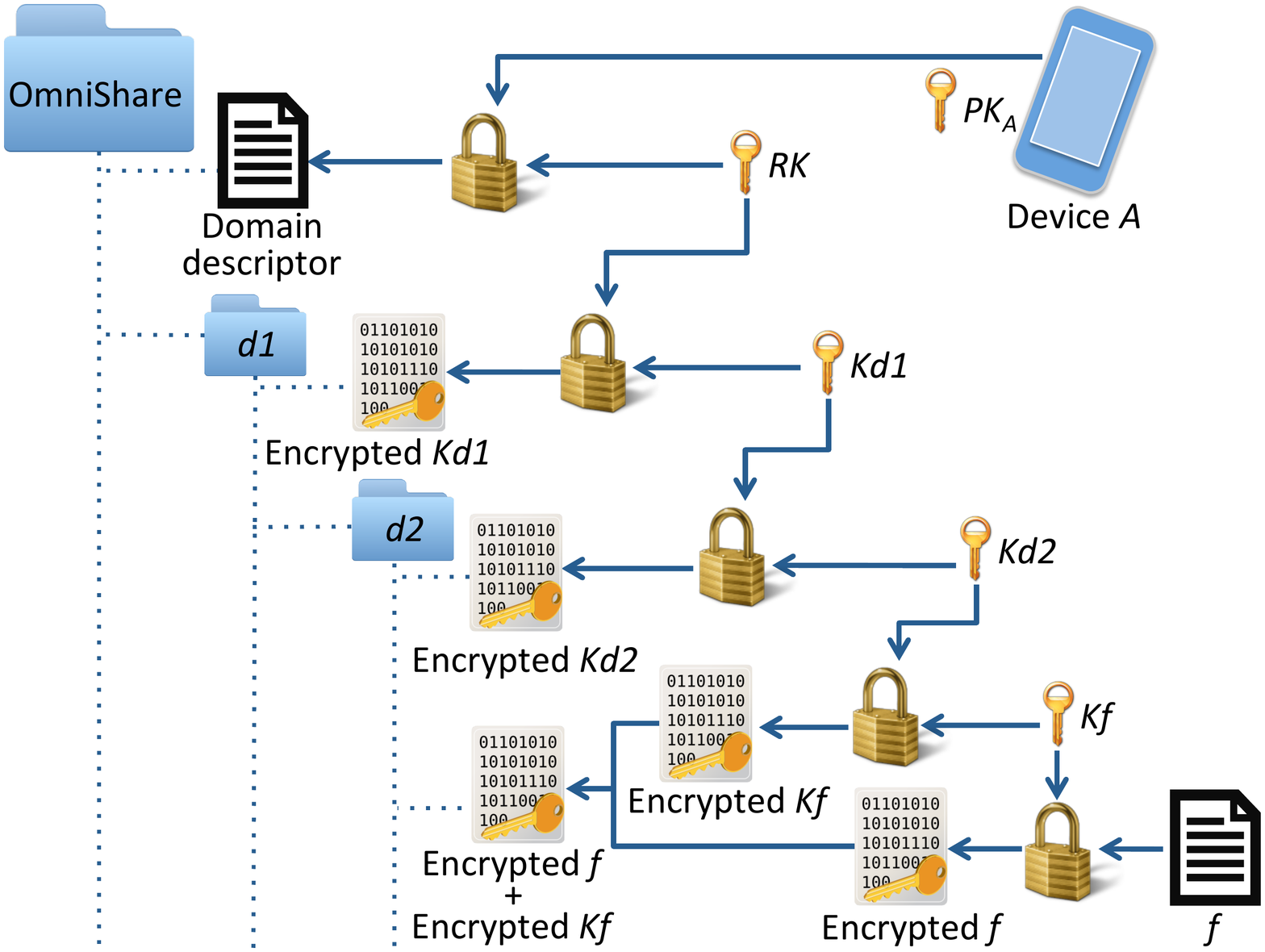}
   \caption{Example of a key hierarchy in \OmniShare}
   \label{fig:keyHierarchy}
\end{figure}

Figure~\ref{fig:keyHierarchy} gives an example of an \OmniShare key hierarchy.
The domain's root key ($\tens{RK}$) is encrypted using the public key of device \textsfsl{A} ($\tens{PK_A}$) and stored in the domain descriptor file.
$\tens{RK}$ is used to encrypt the directory key $\tens{Kd1}$, corresponding to directory \textsfsl{d1}.
This encryption also includes the path of \textsfsl{d1} relative to the \OmniShare root directory.
$\tens{Kd1}$ is in turn used to encrypt directory key $\tens{Kd2}$, since \textsfsl{d2} is a subdirectory of \textsfsl{d1}.
$\tens{Kd2}$ encrypts file key $\tens{Kf}$, which encrypts file \textsfsl{f}.
The encryption of $\tens{Kf}$ also includes a hash of the encrypted file to protect its integrity.
This encrypted $\tens{Kf}$ is stored together with the encrypted file in directory \textsfsl{d2}
whereas $\tens{Kd1}$ and $\tens{Kd2}$ are stored as separate encrypted files in their corresponding
directories.
When an encrypted file is downloaded by an authorized device, \OmniShare decrypts the relevant file key and uses it to decrypt the file on the user's device. 

\subsection{Device Authorization}
\label{sub:devAuth}

\OmniShare uses the same cloud storage service to store both the data and the key hierarchy, thus eliminating the need for additional servers.
Therefore, adding a new device to a domain involves granting this device access to the root key so that it can access all other keys without requiring further interaction (Requirement F2).

\ifsubmission
This device authorization process uses a combination of an out-of-band (OOB) 
communication channel and communication via the \emph{cloud storage} itself.
To exchange messages via the cloud storage channel, devices upload their messages as files with specific names that are recognized as messages by other devices.

\else
\begin{figure}[t]
   \centering
   \includegraphics[trim=2.5cm 1cm 2cm 1cm, clip=true, width=1\linewidth]{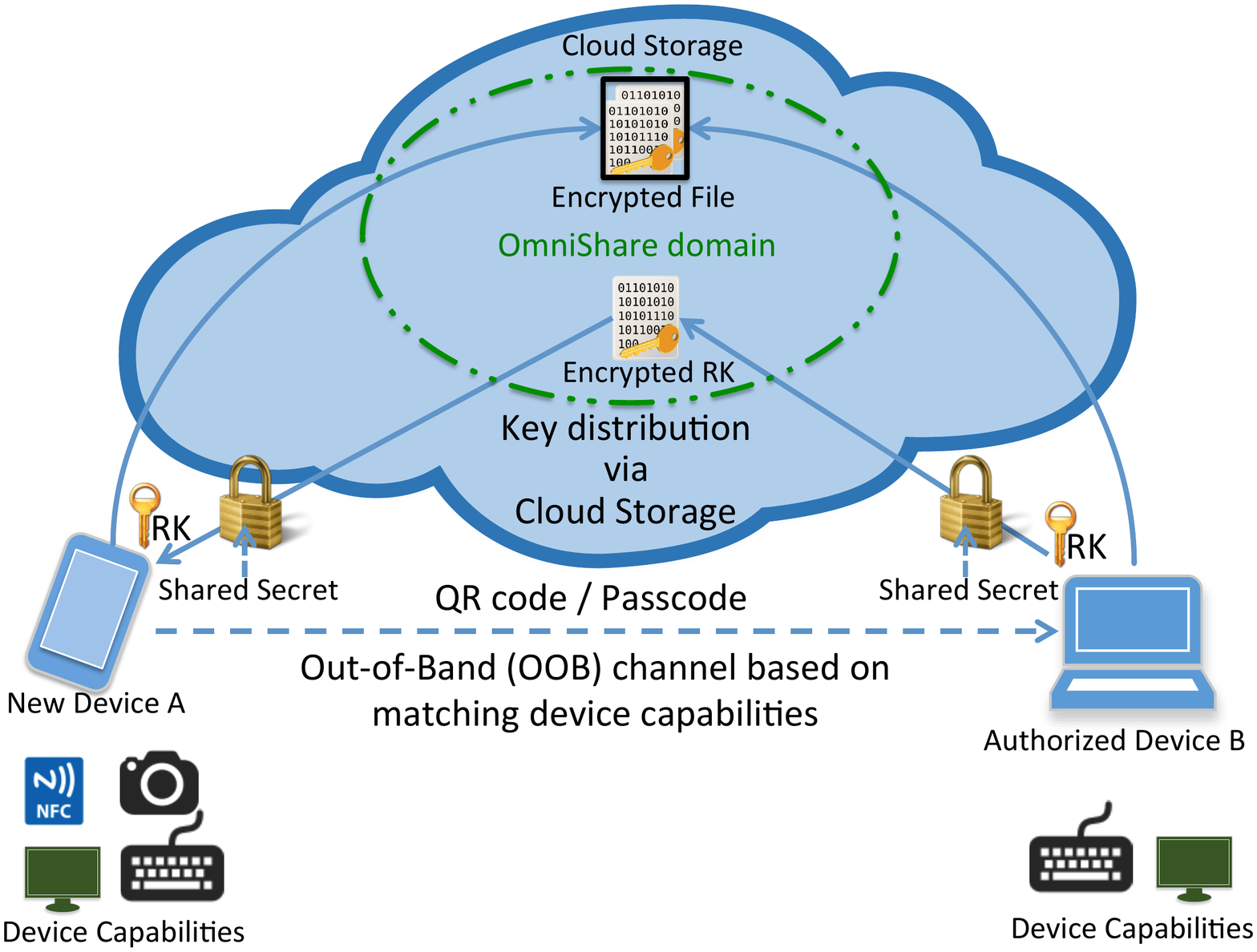}
   \caption{Overview of device authorization in \OmniShare}
   \label{fig:keyDist}
\end{figure}

As shown in Figure~\ref{fig:keyDist}, this device authorization process uses a combination of an out-of-band (OOB) communication channel and communication via the \emph{cloud storage} itself.
To exchange messages via the cloud storage channel, devices upload their messages as files with specific names that are recognized as messages by other devices.
\fi

When a new device (\NewDev) requests to join a domain, \OmniShare allows the user to select a suitable authorizing device (\AuthDev) from within the domain to complete this action.
In a naive approach, device \AuthDev could simply encrypt the root key with device \NewDev's public key.
However, this would not provide any guarantee that the correct device has received the root key (the adversary could have replaced the public key with his own) or that the correct root key has been received (the adversary could have injected his own root key).
To mitigate against these attacks, \OmniShare also uses a low-bandwidth uni-directional OOB channel to authenticate the new device (\NewDev) to the authorizing device (\AuthDev).
Specifically, the OOB channel is used to confirm \NewDev's public key and establish a shared secret between \NewDev and \AuthDev.
For a consistent user experience (Requirement U1), the OOB channel is always a uni-directional transfer of information from \NewDev to \AuthDev.
Although the OOB channel requires only minimal user interaction, this is sufficient to bootstrap the security guarantees for the rest of the system.
Based on the hardware capabilities of the new device and the authorizing device, \OmniShare selects the best type of OOB channel.
Since the OOB channels vary in terms of bandwidth, \OmniShare supports two types of protocols for device authentication: \emph{single round-trip} and \emph{multiple round-trip} protocols, where a \emph{round-trip} refers to an exchange of messages between the devices via the cloud storage communication channel.
Whilst all OOB channels can support the multiple round-trip protocol, certain types of OOB channels can enable the more efficient single round-trip protocol, as explained in the following subsections.

\subsection{Single Round-Trip Protocol}
\label{sub:single-round-trip}

\begin{figure}[t]
   \centering
   \includegraphics[trim=0cm 4.5cm 5cm 0cm, clip=true, width=1\linewidth]{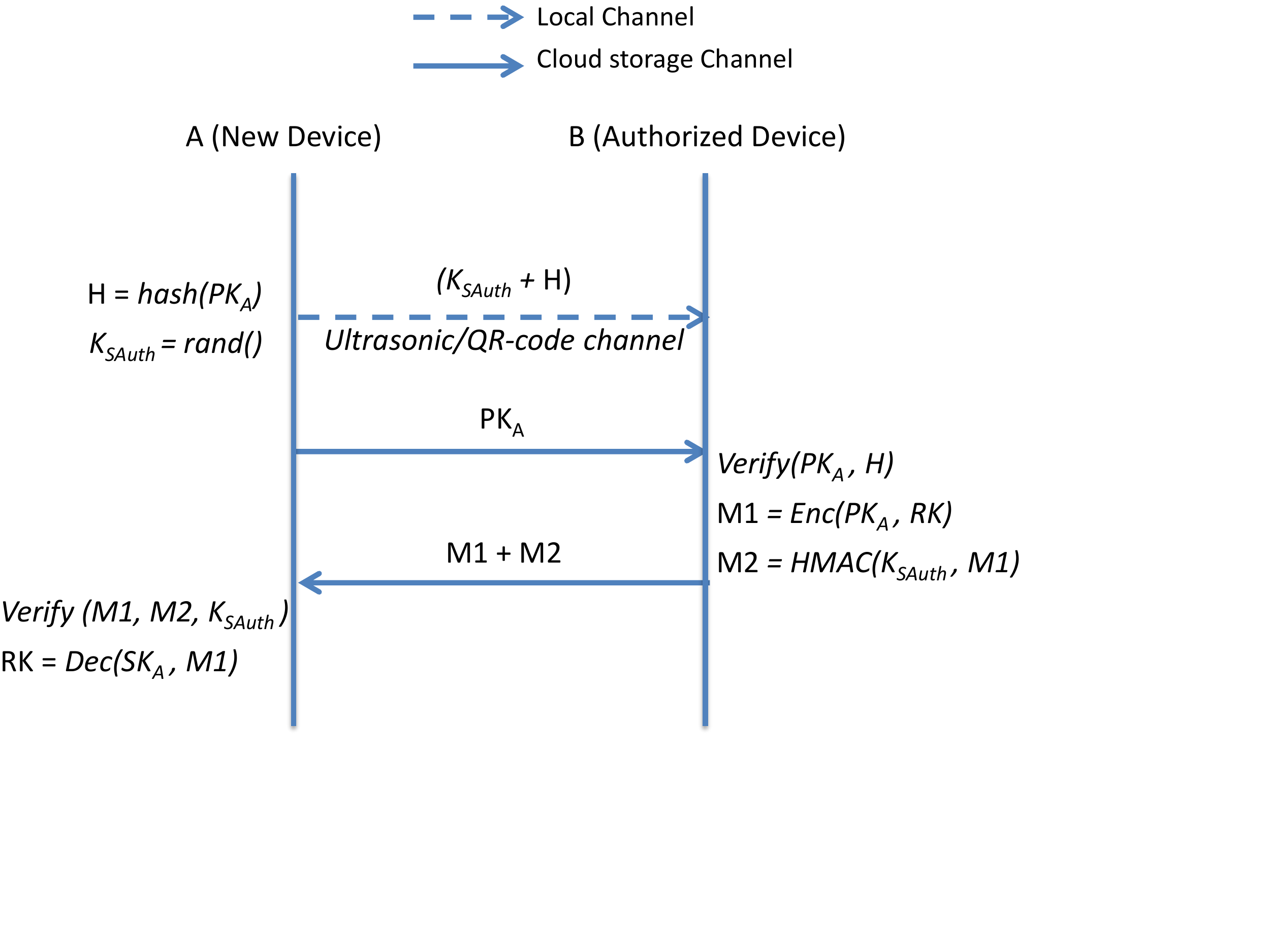}
   \caption{Single round-trip device authorization protocol}
   \label{fig:single-round-trip}
\end{figure}

As shown in Figure~\ref{fig:single-round-trip}, the single round-trip protocol proceeds as follows:

\begin{enumerate}
\renewcommand{\theenumi}{\roman{enumi}}

\item Device \NewDev computes a cryptographic hash $\tens{H}$ of its public key $\tens{PK_A}$ and a freshly-generated random session authentication key $\tens{K_{SAuth}}$ and transfers these to device \AuthDev via the OOB channel.
\ifsubmission
In our implementation, we use a SHA256 hash and a 128-bit authentication key.
\else
\fi
   
\item \NewDev delivers $\tens{PK_A}$ to \AuthDev via the cloud storage channel.

\item After verifying $\tens{PK_A}$, \AuthDev encrypts the root key $\tens{RK}$ with $\tens{PK_A}$. 
\AuthDev also generates a hash-based message authentication code (HMAC) $\tens{M2}$ for the encrypted message $\tens{M1}$ using $\tens{K_{SAuth}}$ as the key.

\item Upon receiving $\tens{M1}$ and $\tens{M2}$ via the cloud storage, \NewDev verifies the authenticity of $\tens{M1}$ and decrypts $\tens{RK}$ from $\tens{M1}$.

\end{enumerate}

This protocol can be used over any OOB channel that provides sufficient bandwidth for device \NewDev to deliver $\tens{H}$ and $\tens{K_{SAuth}}$ to device \AuthDev.
We have implemented the following types of OOB channels.

\revised{
\subsubsection{Ultrasonic Communication}
If the new device (\NewDev) has a speaker and the authorizing device (\AuthDev) has a microphone, the devices can use ultrasonic communication as the OOB channel (i.e. audio frequencies greater than the upper limit of human hearing).
The hash and the authentication key are encoded as a high frequency audio signal, which is played by device \NewDev and recorded by device \AuthDev.
Ultrasonic communication has recently drawn significant interest from both academia and industry~\cite{Li2015, Zhang2014_priwhisper, Santagati2015, Lee2015, Chromecast}. 
It is an ideal OOB channel for \OmniShare because it is range-limited by the physical environment (e.g. by doors and walls) in the same way as a private conversation.
It is also widely deployable due to the prevalence of microphones and loudspeakers in consumer devices.  
\ifsubmission
In our implementation, we use a \emph{chirp signal}~\cite{Cook1974_chirp, Lee2015} to encode each binary bit as either an \emph{up-chirp} (one) or \emph{down-chirp} (zero).
On the receiving side, we decode the received signal by correlating it with the chirp signatures.
We use the open-source \emph{ZXing for Reed-Solomon} library~\cite{ZXingReedSolomon} to provide error-correction codes.
\fi
}

\subsubsection{QR Code}
Similarly, if device \NewDev has a screen and device \AuthDev has a camera, a \emph{QR code} can be used as the OOB channel.
The hash and the authentication key are encoded as a two-dimensional QR code by device \NewDev, which the user scans using device \AuthDev.
\ifsubmission
In our implementation, we use the open-source ZXing~\cite{ZXing} library for Android and Windows to generate and decode 300~x~300 pixel QR codes.
\fi

\subsection{Multiple Round-Trip Protocol}
\label{sub:multiple-round-trip} 

\begin{figure}[th]
   \centering
   \includegraphics[trim=2cm 6cm 6cm 1cm, clip=true, width=1\linewidth]{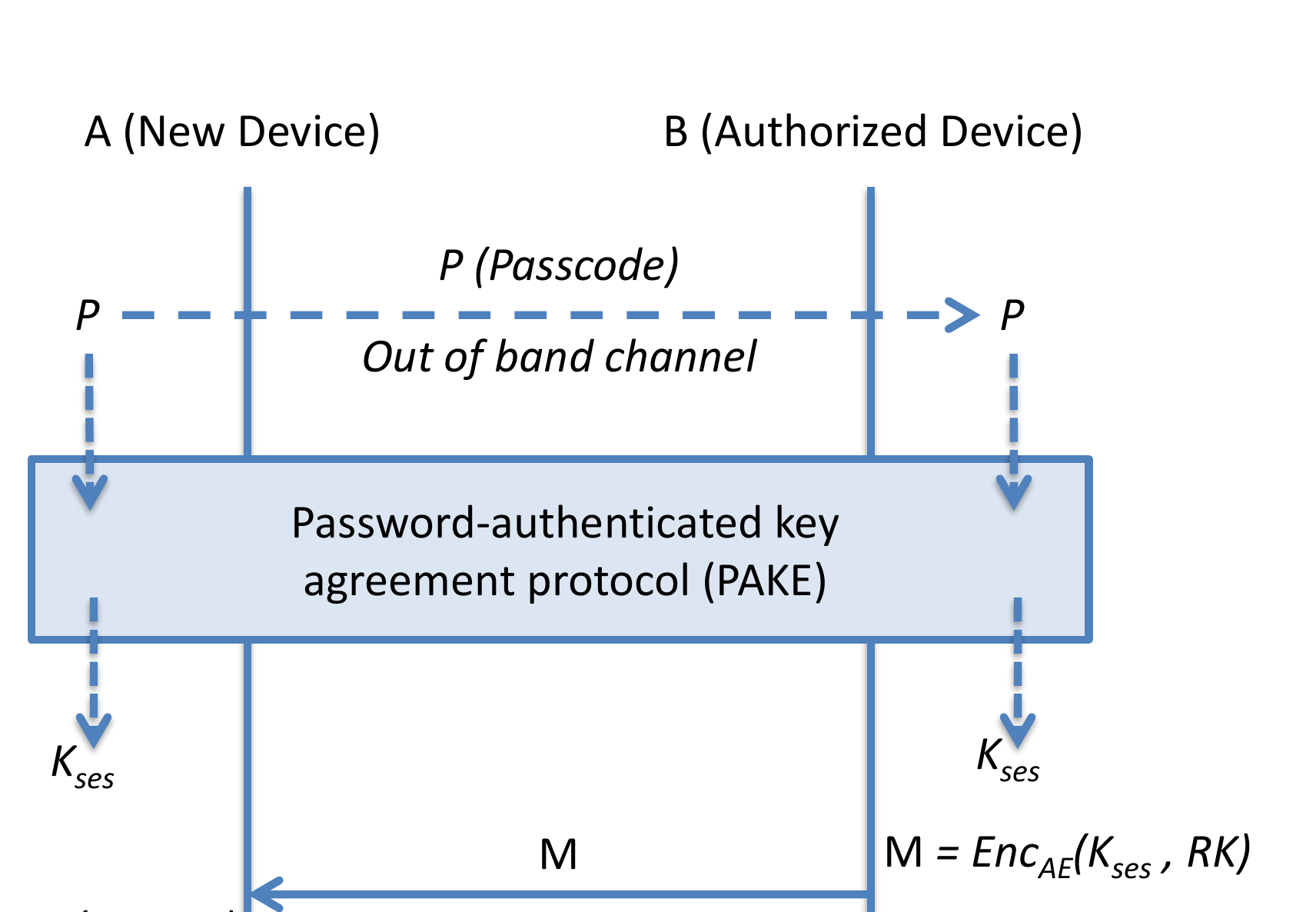}
   \caption{Device authorization protocol using passcode}
   \label{fig:keydistPAKE}
\end{figure}

In this simple device authorization protocol, users enter a passcode displayed on the new device (\NewDev) into the authorizing device (\AuthDev). 
We use a password-authenticated key agreement (PAKE) protocol, to generate a strong shared session key from the shared passcode and use this session key to securely distribute $\tens{RK}$ via the cloud storage.
In our implementation, we use the secure remote password (SRP) protocol version 6a~\cite{SRP,SRP6} to derive a strong 128-bit session key from the 6-digit random passcode.
We chose this particular PAKE variant because it meets all the security requirements and is not encumbered by patents~\cite{SRP}.
\ifsubmission
The full details of this protocol are presented in the accompanying technical report~\cite{arXiv:1511.02119}.
\else
\fi
As shown in Figure~\ref{fig:keydistPAKE}, the overall protocol proceeds as follows:

\begin{enumerate}
\renewcommand{\theenumi}{\roman{enumi}}

\item \NewDev displays a 6-digit passcode $\tens{P}$ which the user types into \AuthDev via its input keyboard.

\item Both devices run a PAKE protocol via the cloud storage and derive a shared session key $\tens{K_{ses}}$.

\item \AuthDev encrypts $\tens{RK}$ using an authenticated encryption algorithm $\tens{Enc_{AE}(K_{ses}, RK)}$ with $\tens{K_{ses}}$ and delivers the encrypted message $\tens{M}$ to \NewDev via the cloud storage.

\item \NewDev decrypts $\tens{M}$ using the corresponding decryption algorithm $\tens{Dec_{AE}(K_{ses}, M)}$ with $\tens{K_{ses}}$ to extract $\tens{RK}$.

\end{enumerate}

\ifsubmission

\else
\begin{figure}[th]
   \centering
   \includegraphics[trim=2cm 1cm 6cm 0cm, clip=true, width=1\linewidth]{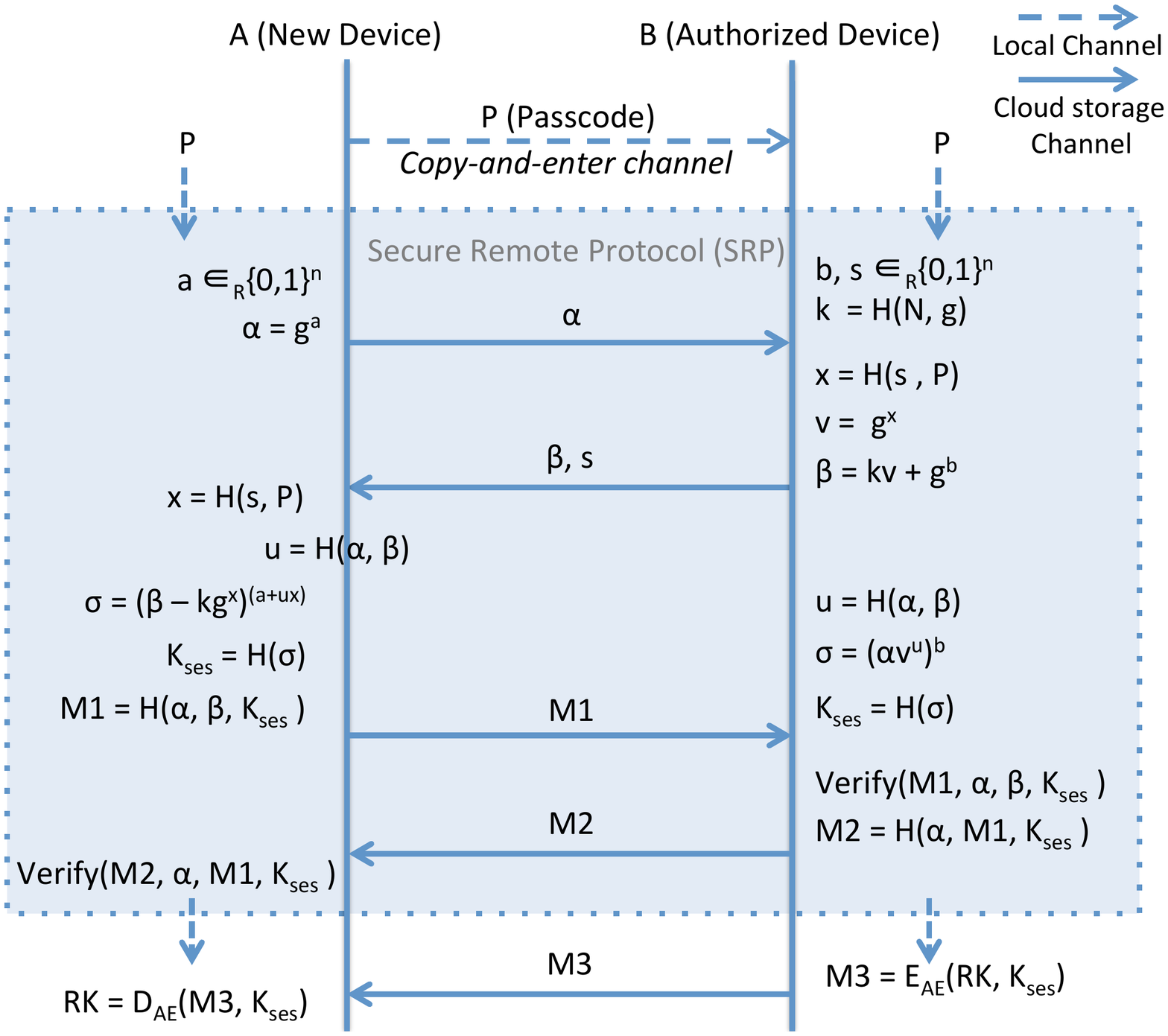}
   \caption{Device authorization protocol using passcode}
   \label{fig:keydistSRP}
\end{figure}

The secure remote password (SRP) protocol~\cite{SRP6} is used as the PAKE
protocol to derive $\tens{K_{ses}}$ from $\tens{P}$. SRP is a 
client-server protocol in which clients first register with the server and set up their passwords.
The server generates random salts and stores the cryptographic hashes
of the clients' salted passwords. During authentication, clients
provide provide their identities to the server and request the respective salt values. 


Figure~\ref{fig:keydistSRP} outlines the device authorization protocol using a
passcode where \NewDev is the SRP client and \AuthDev is the SRP
server. Both devices use a 6-digit numeric passcode $\tens{P}$ as the SRP password. However, we
omit the registration phase and the first SRP message (i.e. sending the
client's identity) since \OmniShare has only one client and
server during authorization. SRP uses a finite field $\tens{GF(N)}$ for all
computations where $\tens{N}$ is a large prime and $\tens{g}$ is a generator
in $\tens{GF(N)}$. Both devices use same $\tens{N}$ and $\tens{g}$. The protocol is as follows:

\begin{enumerate}
\renewcommand{\theenumi}{\roman{enumi}}
	 \item \NewDev generates a random number $\tens{a}$, calculates its public value 
	 $\tens{\alpha = g^a}$ and transfers $\tens{\alpha}$ to \AuthDev. Meanwhile, \AuthDev
	 generates a random number $\tens{b}$, a random salt $\tens{s}$. Both calculate the cryptographic hash
	 $\tens{k = H(N, g)}$.
	 
	 \item After receiving $\tens{\alpha}$, \AuthDev calculates $\tens{x = H(s, P)}$, $\tens{v = g^x}$ and its public value $\tens{\beta = kv + g^b}$. \AuthDev then transfers $\tens{\beta}$ and $\tens{s}$ to \NewDev.

	 \item \NewDev calculates the cryptographic hash $\tens{x = H(s, P)}$. Both \NewDev and \AuthDev calculate $\tens{u = H(\alpha, \beta)}$ and the common value $\tens{\sigma} = \tens{(\beta - kg^x)^{(a + ux)}} = \tens{(\alpha v^u)^{b}}$. Both devices then hash $\tens{\sigma}$ to derive a session key $\tens{K_{ses}}$. After calculating $\tens{K_{ses}}$, \NewDev calculates the session key authentication message $\tens{M1 = H(\alpha, \beta, K_{ses})}$ and delivers $\tens{M1}$ to \AuthDev.

	 \item \AuthDev verifies $\tens{M1}$ using $\tens{verify(M1, \alpha, \beta, K_{ses})}$ and sends its part of the session key authentication message $\tens{M2 = H(\alpha, M1, K_{ses})}$ to \NewDev. 
\end{enumerate}

After completing the SRP protocol, \AuthDev uses the $\tens{K_{ses}}$ to securely deliver the $\tens{RK}$ to \NewDev as explained above.

\fi

Once authorized, the \OmniShare client on \NewDev adds \NewDev to the user's \OmniShare domain by performing the following tasks: (a) encrypting the $\tens{RK}$ with $\tens{PK_A}$, (b) calculating the HMAC of the device metadata and the encrypted $\tens{RK}$ with the device-specific authentication key, and (c) adding the HMAC along with the encrypted $\tens{RK}$ and the device metadata into the domain descriptor file.

\subsection{Sharing Files}
\label{sub:filesharing}

\OmniShare supports sharing selected files with other users. 
Since the system is not limited to any particular cloud storage provider (Requirement~F4), we cannot assume that collaborative sharing capabilities (i.e. concurrent editing of the same file by multiple users) will be available.
We therefore provide a read-only sharing mechanism, which can be used with any cloud storage provider.
By sharing an encrypted file and the corresponding individual file key, the receiver can read the file but cannot make modifications without causing the integrity check (the hash in the encryption of $\tens{Kf}$) to fail.
File sharing is efficient in the sense that files do not need to be re-encrypted in order to be shared securely (since re-encrypting large files may take a long time).
Although the sharing permission is inherently delegable, this is no different from users passing on the contents of shared files.
The sharing arrangement can be terminated by re-encrypting the files with new keys.
Files can also be shared with groups of users by distributing the relevant keys to multiple receivers.

Specifically, file sharing involves three main tasks: \textit{Peering}, \textit{Sharing} and \textit{Receiving}. There is also an optional task \textit{Storing}.

\subsubsection{Peering}
When two users want to share files, they first run a key exchange protocol over an OOB channel, to agree on a shared \textit{peer key}. 
\ifsubmission
In our implementation we use NFC and Bluetooth to establish bidirectional communication between a pair of devices, and perform an Elliptic Curve Diffie-Hellman (ECDH) key exchange to derive a 256-bit shared key.
\else
\fi
Each peer then establishes a persistent context for the peering consisting of this shared key, along with a \textit{peer directory} and a \textit{control file}. Each device sends a public link to its control file to the peer. Each device maintains a list of added peers and pointers to their control files. 

\subsubsection{Sharing}
When the user shares a file with a receiver, \OmniShare first copies the encrypted file(s) to the peer directory for the receiver and adds a record to the control file in the peer context containing: (a) the file or directory key encrypted with the peer key and (b) a public link to the encrypted file or directory. 

\subsubsection{Receiving}
When the receiver scans the control file of a peer, it can detect all newly shared files from that
peer. The receiver can use the public links to fetch the encrypted files. It can also fetch the encrypted keys from the peer's control file. Using these and the corresponding peer key, the receiver can recover the files' contents. On successfully receiving the files, the receiver adds an acknowledgement entry, i.e. the identities of the files, in her control file as an acknowledgement to the sender. 

\subsubsection{Storing}
This is an optional task in the sharing process in which the receiver imports the shared file(s) into her cloud storage using the key hierarchy of her own \OmniShare domain.

\revised{
\subsection{Extensions}

In addition to the core architecture described above, the following features could naturally be integrated into \OmniShare:

\textbf{Selective synchronization:}
In certain circumstances, a user may wish to synchronize only a subset of her files to one of her devices.
The \OmniShare key hierarchy is well suited for this purpose.
Due to the symmetry of keys, a new device can be given a directory key in place of the root key, thus authorizing access to only a subset of the directory tree.

\textbf{Directory sharing:} 
The same mechanism used for sharing files with other users could also be used to share directories, by replacing the file keys with directory keys.

\textbf{Server-side computation:} 
It might be argued that client-side encryption limits the possibility for honest cloud providers to perform computations on the encrypted files (e.g. search and analysis).
However, new types of encryption schemes, such as order-preserving encryption (OPE)~\cite{Agrawal2004} and fully-homomorphic encryption (FHE)~\cite{Gentry2009}, which allow providers to perform some types of computations directly on the encrypted files, could be used in \OmniShare.

\textbf{Delta file encryption:} Naively, updating an encrypted file involves re-encrypting the entire file, which may be expensive in terms of processing and bandwidth, especially for large files.
Instead, changes could be recorded as separate encrypted \emph{delta files}, which are also synchronized across client devices. 
When decrypting files, client devices also decrypt the associated delta files and apply the changes locally.
Delta file include an integrity-protected \emph{last modified} timestamp to prevent rollback attacks.  
The cloud provider can also use this mechanism to enable deduplication of encrypted files by storing a single copy of similar files and maintaining the differences using delta files.
}


\ifsubmission


\else

\section{Implementation}
\label{sec:implementation}

We have implemented \OmniShare on Windows and Android, and support Dropbox as the cloud storage service.
However, support for other platforms and cloud storage providers can be added without modifications to the architecture (Requirement F4). 

On Windows, the implementation uses the .NET framework (version 3.5) 
for x86 and x64 architectures. The Android implementation
targets Android 4.1 and higher (API level 16). Both platforms use the Bouncy Castle Crypto
APIs~\cite{BouncyCastle} (Bouncy Castle C\#
v1.7 on Windows and Spongy Castle v1.51 on Android). A
port for iOS 7.0 and higher is in progress.
We have implemented \OmniShare for Dropbox but the architecture is not limited 
to this provider (Requirement F4). Adding support for other providers is straightforward 
provided they offer interfaces for third-party applications.

\subsection{File Encryption and Key Hierarchy}
\label{sub:implementation-encryption}

We use the Advanced Encryption Standard in Galois Counter Mode (AES-GCM) with a 128-bit key to encrypt files and keys. 
Since the cloud storage provider may be colluding with the adversary, this semantically secure encryption is used to prevent the adversary from learning any information about the file's contents.

\begin{figure}[h]
   \centering
   \includegraphics[trim=2cm 17cm 8cm 1cm, clip=true, width=1\linewidth]{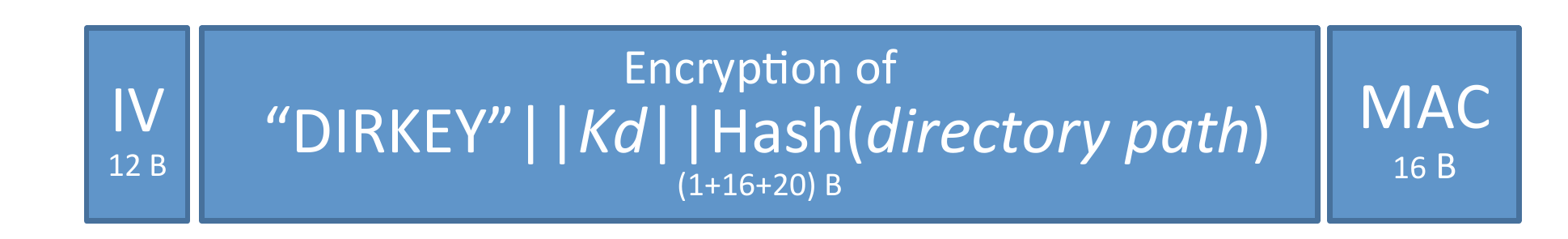}
   \caption{Format of a .omnishare.envelope file (i.e. an encrypted \folder key) }
   \label{fig:envelope}
\end{figure}

\Folder keys are encrypted together with a 1-byte tag indicating the key type (shown as the constant \textit{DIRKEY}) and the hash of their full \folder paths in order to mitigate against key-substitution attacks within the same key hierarchy.
Encrypted \folder keys are stored as \textit{.omnishare.envelope} files in
their corresponding \folders. Figure~\ref{fig:envelope} shows the format of a
\textit{.omnishare.envelope} file where the IV is prepended to the encryption
of the \folder key $\tens{Kd}$ and the MAC is appended after the encrypted key.
Similarly, each encrypted file key also includes a 1-byte key type tag
(e.g. \textit{FILEKEY}) and the hash of the encryption of the
corresponding file. Figure~\ref{fig:encryptedkey} shows the format of an
encrypted file key where the IV is prepended to the encryption of the file key
$\tens{Kf}$ and the MAC is appended to the encryption. The encrypted file keys
are prepended to the encryption of the corresponding files (although note that 
this is purely for convenience and does not provide any security properties 
given our adversary model). The encryption of
the files also follows a similar format as the keys with their IVs prepended and
MACs appended after the encryption. However, the encrypted files do not include 
the additional information included with the encrypted keys.

\begin{figure}[h]
   \centering
   \includegraphics[trim=2cm 17cm 8cm 1cm, clip=true, width=1\linewidth]{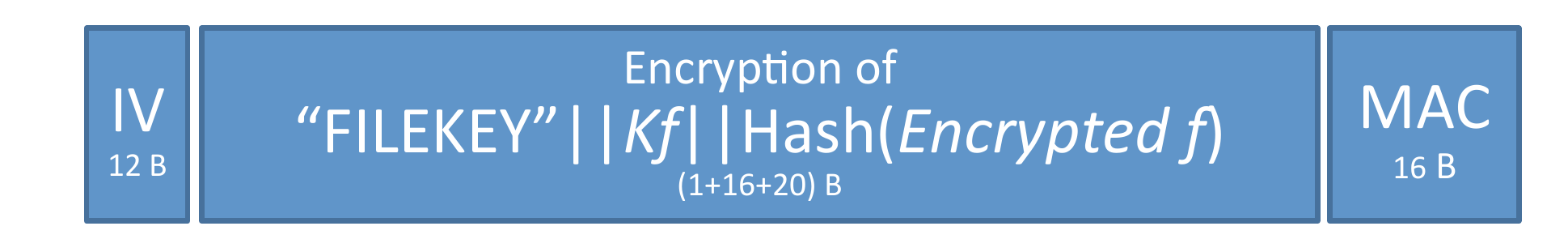}
   \caption{Format of an encrypted file key}
   \label{fig:encryptedkey}
\end{figure}



\subsection{Device Authorization}

Protocol messages are exchanged as files via the cloud storage service itself.
Each message file has a Universally Unique Identifier (UUID) as a filename and contains a JSON object with the protocol message.
When a response is expected, the UUID filename for the response file is also specified. 
The filename of the initial authorization protocol message indicates the identity of the selected authorizing device, so that only this device responds.
However, this naming convention is purely for convenience since device authenticity and message freshness are both established cryptographically during the protocol.


\subsubsection{Ultrasonic Communication}
When using ultrasonic OOB channel, we calculate the SHA256 hash of the device public key and encode this together with a 128~bit random authentication key as an ultrasonic audio signal. 
We use a \emph{chirp signal}~\cite{Cook1974_chirp, Lee2015} to encode each binary bit as either an \emph{up-chirp} (one) or \emph{down-chirp} (zero).
Through experimentation with different devices, we determined that the optimal frequency band is from 16.2~kHz to 17.2~kHz, and we achieved a bitrate of 100~bps. 
On the receiving side, we decode the received signal by correlating it with the chirp signatures.
We use the open-source \emph{ZXing for Reed-Solomon} library~\cite{ZXingReedSolomon} to provide error-correction codes.

\subsubsection{QR Code}
Similarly, when using the QR~Code OOB channel, the SHA256 hash of the device public key and the 128~bit authentication key are encoded as a QR code. 
We use the open-source ZXing ("zebra crossing")~\cite{ZXing} library for Android and Windows to generate and decode 300~x~300 pixel QR codes.

\subsubsection{Passcode}
When using the Passcode OOB channel, we use the secure remote password (SRP) protocol version 6a~\cite{rfc2945} as the PAKE protocol to derive a strong 128-bit session key from the 6-digit random passcode.
We use this session key to distribute the root key.

\subsection{Sharing Files}

The capability to share files with other users is only implemented in the Android version.
We use ultrasonic communication, NFC, or Bluetooth to establish bidirectional communication between a pair of devices.
The devices perform an Elliptic Curve Diffie-Hellman (ECDH) key exchange to derive a 256-bit AES shared key. 
The cloud storage provider's API (e.g. the Dropbox API) is used to generate public links of the control files and shared files. 
The current implementation does not yet support sharing directories or selectively synchronizing directories, although these features are under development.

\fi

\section{Evaluation}
\label{sec:evaluation}


We evaluate the security, functionality and usability of \OmniShare in terms of the requirements defined in Section~\ref{sec:requirements} and we benchmark the performance of our Android and Windows implementations.

%

\subsection{Architecture Security Evaluation}
\label{sub:evaluation-architecture}

\OmniShare uses standardized cryptographic algorithms and high entropy keys for all cryptographic operations, thus fulfilling Requirement~S1.
Assuming the adversary is unable to subvert these cryptographic algorithms, he cannot read the encrypted files or keys in the hierarchy without access to the root key. 
Since the root key itself is encrypted with the public keys of authorized devices, an adversary outside the domain cannot access this key, thus fulfilling Requirement~S2.
Furthermore, since \OmniShare uses authenticated encryption and HMACs, any unauthorized modification of files or substitution of encrypted keys can be detected.
At most, the adversary can learn the filenames and approximate file sizes (the exact file sizes are masked by the encryption padding).
We have chosen not to encrypt the filenames so that the cloud provider can offer filename-based search and so that mobile clients can selectively download and decrypt files in order to reduce bandwidth and energy consumption.

\revised{
Our use of a key hierarchy and individual file keys facilitates efficient sharing and limits the consequences if certain keys are compromised.
When sharing a file (i.e. read-only sharing), the user reveals only a single file key to the receiver (who may collude with the adversary).
Although this gives the adversary the ability to read that specific file, the file can only be modified by the original user since the file's integrity-check value is protected by the corresponding directory key.
Due to the key hierarchy, \OmniShare does not always need to keep the root key in memory.
For example, when the user is working in a specific directory, the root key can be only used briefly to decrypt the corresponding directory key.
In general, we assume that the adversary cannot read the memory of the \OmniShare application on a legitimate user's device.
However, in a real-world deployment, there may still be attacks through which an adversary could extract secrets from memory (e.g. a \emph{cold boot attack}). 
Thus removing the root key from memory when it is not in use is a \emph{defence in depth} mechanism that reduces the window of vulnerability during which this high-value key is in memory.

The consequences of specific key compromises are also reduced.
If a file key is compromised, only a single file is vulnerable.
Similarly, compromise of a directory key only exposes a single directory.
Either of these two types of compromise only require re-encryption of the affected files.
In the worst case scenario, compromise of the root key would expose all files.
However, since the root key is protected by the device keypair using asymmetric cryptography, additional mechanisms can be used to protect the device keypair.
For example, on Android, the device's private key is protected by the Android keystore, which is often backed by secure hardware, making it very difficult for malware to extract this key.
Furthermore, the keystore can require user authentication before allowing access to this key, thus protecting this key if the device is stolen.
}

\subsection{Protocol Security Evaluation}
\label{sub:evaluation-protocol}

In addition to the architecture itself, we evaluate the security of our proposed device authorization protocols using the Scyther protocol analysis tool~\cite{scyther}.
Specifically, for both the single and multiple round-trip protocols, we analyse the following security properties:
\begin{itemize}
\item Once the protocol is complete, the new device and the authorizing device will both have access to the same root key (agreement property).
\item The adversary cannot learn this root key unless it is explicitly authorized by the user (secrecy property).
\end{itemize}
\ifsubmission
The full formal models of our protocols are presented in the accompanying technical report~\cite{arXiv:1511.02119}.
\else
An overview of the Scyther tool is given in Appendix~\ref{sec:scyther} and the full formal models of our protocols are presented in Appendix~\ref{sec:scyther-single} and Appendix~\ref{sec:scyther-multiple}.
\fi
The analyses confirm that these properties hold for both protocols with respect to an adversary who has full control over the network and cloud storage but cannot interfere with the OOB channels.
As explained in Section~\ref{sec:requirements}, these are realistic assumptions of the adversary's capabilities.
Finally, the user's involvement in the OOB channel, although minimal, is sufficient to authorize the transaction and bootstrap trust between the two devices.
Therefore, both protocols fulfil Requirement~S2.
Although the security of the PAKE protocol that forms the basis of \OmniShare's multiple-round-trip protocol has also been evaluated using cryptographic proofs, these proofs only deal with the correctness of the protocol (i.e. an adversary without the password cannot complete the protocol) and the confidentiality of the password.
However, since these cryptographic proofs consider PAKE protocol in isolation, they cannot reason about other properties, such as authentication between communicating entities, that emerge when the PAKE protocol is used as part of a larger system.
Therefore, in addition to the cryptographic proofs, it is essential to analyse the PAKE protocol in context, using symbolic analysis tools, such as Scyther.

\subsection{Performance Evaluation}
\label{sub:evaluation-performance}

\ifsubmission 

We benchmarked our implementation using a Microsoft Surface Pro (Windows~10, Core~i5~1.7~GHZ) and a Samsung Galaxy S6 (Android 5.1, Quad-core 1.5~GHz Cortex-A53 + Quad-core 2.1~GHz Cortex-A57).
In our Windows implementation, the 2048~bit RSA encryption of the root key (16~kB) required \textless 1~ms while decryption took 13.0~ms.
Encrypting a 1~MB file with a 128~bit AES key required 277.0~ms and decrypting this file took 265.5~ms.
Our Android implementation achieved similar performance in same scenarios requiring 15.2~ms or 31.4~ms for RSA encryption or decryption, and 211.6~ms or 235.0~ms for AES encryption or decryption.
Therefore, although file encryption and decryption are computationally intensive operations, our benchmarks show that these can still be achieved in reasonable time for this application.

We also measured the average latency of our two device authorization protocols.
The average latency of the QR~code protocol was 18.99~seconds and of the Passcode protocol was 40.64~seconds (excluding any additional time taken by the user to complete the OOB channel).
This difference arises because the QR~code protocol requires half the number of messages of the Passcode protocol.
However, since these protocols use the cloud storage service as a communication channel, this latency is almost entirely due to the latency of the cloud service in synchronizing files.
If the latency of the cloud service decreased, this would significantly decrease the latency of our protocols.

\else

We used a Microsoft Surface Pro running Windows 10 with a core~i5~1.7~GHZ CPU and a Samsung Galaxy S6 running Android 5.1 using both a Quad-core 1.5~GHz Cortex-A53 and a Quad-core 2.1~GHz Cortex-A57 CPU for our measurements. 
The values below are the average of 10 execution rounds for each measurement.

\begin{table}[ht]
\centering
\caption{Measurements of device authorization protocol execution time using Dropbox}
\label{table:measurement}
\begin{tabular}{l r r r}
\hline\hline

Authorization protocol 	 & \multicolumn{3}{r}{Average time (seconds)}\\[0.5ex] 
\hline 
\textbf{Single round-trip}\\[0.5ex] 
Windows $\rightarrow$ Android  && 16.31 & ($\pm$ 2.37)  \\
Android $\rightarrow$ Android && 21.66 & ($\pm$ 1.10)  \\
\hline
\textbf{Using SRP}\\[0.5ex] 
Windows $\rightarrow$ Android && 39.77 & ($\pm$ 4.08) \\
Android $\rightarrow$ Windows && 36.68 & ($\pm$ 9.21) \\
Windows $\rightarrow$ Windows && 45.10 & ($\pm$ 5.71) \\
Android $\rightarrow$ Android && 41.01 & ($\pm$ 3.94) \\ 
\hline
\\
\end{tabular}
\end{table}

Table \ref{table:measurement} shows execution time for the device authorization protocols. In the table Windows $\rightarrow$ Android indicates that the new device is a Windows PC and the selected authorized device is an Android phone.
The measurement includes message generation and exchange time via cloud storage but does not include time for user interaction over OOB channels.
As expected, the table confirms that the single round-trip authorization protocol is at least twice as fast
as as the multiple round-trip protocol. 
On average it takes 16271 ($\pm$19.21) milliseconds to perform one request-response operation between two devices via the Dropbox Sync API.
This time can be reduced by using the Dropbox Core API but this requires a custom file synchronization mechanism.

\begin{table}[ht]
\centering
\caption{Measurements of cryptographic operations}
\label{table:crypto}
\begin{tabular}{l r r r}
\hline\hline
Operation   &  \multicolumn{3}{r}{Average time (milliseconds)}\\[0.5ex] 
\hline 
\textbf{Windows}	\\[0.5ex] 
2048-bit RSA keygen && 93.0 & ($\pm$ 34.5)   \\
RSA encryption (16 bytes RK)    && \textless 1 & \\
RSA decryption (16 bytes RK)    && 13.0 & ($\pm$ 1.4)\\
File encryption (1MB)   && 277.0 & ($\pm$ 19.5) \\
File decryption (1MB)   && 265.5 & ($\pm$ 14.3) \\                             
\hline
\textbf{Android} \\[0.5ex] 
2048-bit RSA keygen 			&& 395.6 & ($\pm$ 184) \\
RSA encryption (16 bytes RK)    && 15.2  & ($\pm$ 3.96)   \\
RSA decryption (16 bytes RK)    && 31.4  & ($\pm$ 2.79)  \\
File encryption (1MB)   && 211.6 & ($\pm$ 16.27) \\
File decryption (1MB)   && 235.0 & ($\pm$ 4.47) \\[0.5ex] 
\hline
\\
\end{tabular}
\end{table}

Table \ref{table:crypto} shows the time required to generate a device RSA key pair, encrypt/decrypt
the 128-bit root key using RSA and symmetrically encrypt/decrypt a
1~MB file using AES-GCM. Other operations, such as generating an AES 128-bit key or
encrypting \filekey using AES, take less than a millisecond on both platforms.
Although file encryption and decryption are computationally intensive operations, our benchmarks show that these can still be achieved in reasonable time for files up to a few megabytes in size.

\fi

%

\revised{
\subsection{Usability Evaluation}
\label{sub:evaluation-usability}

We conducted a cognitive walkthrough to evaluate the usability of \OmniShare. A cognitive walkthrough is a long-standing methodology for usability evaluation where system tasks are inspected in detail, and potential usability problems are evaluated for every task~\cite{Wharton.1994}. They are particularly used in situations where the interest is in an in-depth examination of one system, rather than a comparison of multiple systems. 
Cognitive walkthrough is one of the most prominent usability evaluation methods~\cite{Nurseetal2011}, and has been applied in many security-related domains \cite{Clarketal2007}\cite{Xunetal2008}\cite{BennettStephens2009}.
We chose it for its combination of practical feasibility and attention to detail. The purpose of the walkthrough is to uncover potential usability issues a naive first-time user may encounter, by focusing on learnability considerations and explicitly acknowledging the acquisition of skills required to use a system. The full details of this usability evaluation are presented in the accompanying technical report~\cite{} and we summarize our methodology and salient results in this section.

\ifsubmission 
\else 


Security systems have a number of distinct properties that affect their usability~\cite{WhittenTygar1999}. Unlike traditional user interfaces, security systems must be designed so that users cannot make dangerous errors, and must not leak important information to attackers while still providing sufficient feedback to legitimate users. In addition to asking the standard cognitive walkthrough evaluation questions at each step (\emph{Will the user know what to do?}, \emph{Will the user see how to do it?}, and after they have completed the action, \emph{Will the user know that they did the right thing?}), we also paid specific attention to preventing security errors, and minimizing information leakage. 


\fi

\subsubsection{Method}

\ifsubmission

We chose a pluralistic walkthrough with five evaluators in order to represent a spectrum of expertise in our evaluation. Our evaluators included \OmniShare developers, independent usability experts, and a naive first-time user of the system.

We constructed a scenario of prototypical \OmniShare use that included four tasks: setting up \OmniShare for first time use, uploading a file to the \OmniShare directory, accessing the directory from another device, and sharing a file with another user. These tasks represent the functionality available in \OmniShare, and require distinct actions for completion. We subdivided each task into minimal subtasks and evaluated the actions needed to complete, and the potential usability issues arising in each. The evaluation took approximately three hours to complete.

Our evaluation focus was on learnability, and we made minimal assumptions about a user's prior knowledge. We assumed basic competence with the devices used in the walkthrough (a smartphone and a Windows computer), and that the user had previous experience with cloud services, but no technical background. We considered usability problems to be cases where the emulated user could not have completed the task with this minimal background.

\else 

We constructed a scenario of prototypical \OmniShare use that included four tasks. These were (1) setting up \OmniShare for first time use, (2) uploading a file to the \OmniShare directory, (3) accessing the directory from another device, and (4) sharing a file with another \OmniShare user. These tasks were chosen to represent the functionality available in \OmniShare, and required distinct actions for completion. The focus of our evaluation was on learnability and we made minimal assumptions about users' prior knowledge. We assumed basic competence with the devices used in the walkthrough (a smartphone and a Windows computer), and we assumed that the user had previous experience with cloud services, but no technical background.

We chose a pluralistic walkthrough with five evaluators in order to represent multiple viewpoints in our evaluation. Of the five participants, two were directly involved in \OmniShare's design and development, two were independent usability experts, and the final participant was an outside user with no relevant background who represented a naive user.

\fi

\subsubsection{Results}

The overall result of our cognitive walkthrough was that the design of \OmniShare should not present any major usability problems for a users with minimal knowledge of file sharing and little technical background. We concluded that sharing files in \OmniShare should be straightforward for a novice user; although we noted several places in which the user interface could clarify instructions, or align better with operating system standards, we found that a novice user would be able to easily avoid major errors. As part of our cognitive walkthrough, we also uncovered two conceptual issues affecting the design of \OmniShare. These are fundamental issues that arise in file sharing systems, and as such, are not necessarily specific to \OmniShare's architecture (though they affect it).






\ifsubmission

\else

Overall, we found that a naive user would likely be able to complete all tasks in \OmniShare. However, we also identified places in all four tasks where improving language and feedback would enhance the user experience and minimize the possibility of errors. We found instances where language used was either too technical (e.g., ``Access rights''), or where it was ambiguous. We also found problems relating to the placement and labelling of buttons. For example, in the upload task, the file upload interface gave no buttons allowing the user to navigate through the file hierarchy. In all tasks, we found that additional feedback was needed to help the user understand that the tasks had been successfully completed, and what they should do next. This particularly affected the upload and device-pairing tasks. 



We found one dangerous error in our cognitive walkthrough. In the interface, the menu item to disconnect the device from the \OmniShare account was located next to the button to upload a file. Uploading a file is an action that users will need frequently, but disconnecting the device permanently deletes the encryption keys stored on the device and is an irreversible action. If the disconnected device is the only device associated with the account, access to the files in the \OmniShare folder will be permanently lost. This error was corrected by removing the upload file button from that menu and placing it as a stand-alone button on the main screen.

\fi

\ifsubmission
The first of these issues is the nature of how files are shared in \OmniShare. Rather than having a file that is concurrently accessible by multiple people (as in other cloud file-storage systems, such as Dropbox), sharing in \OmniShare is more akin to \emph{sending} a file to another user. In this way, the term ``share'' is used in its active sense, and we suggest that using a different word (perhaps \emph{send} or \emph{transfer}) could help users build better mental models of what is happening when they share a file with another user.

\else 

The final task in the cognitive walkthrough was to share a file with another \OmniShare user for the first time. To do this, the devices need to be paired and users must meet in person to pair their devices (after this step, files can be shared at any time and users do not need to be co-located). Requiring users to pair in person has security advantages, but disadvantages to usability. The larger issue that became apparent in this task was the nature of how files are shared in \OmniShare. Rather than having a file that is accessible by multiple people concurrently (as in other cloud file-storage systems, such as Dropbox), sharing in \OmniShare is more akin to \emph{sending} a file to another user. In this way, the term ``share'' is used in its active sense.
However, this latter interpretation bears some resemblance to the meaning of the term ``share'' in a social media context. One possible way of addressing this issue might be to use a different word (e.g. \emph{send} or \emph{transfer}) in order to help users build a better mental model of the underlying process.

\fi

The other high level issue uncovered by the evaluation was that the interface currently gives the user only mimimal information about encryption of the files. In particular, the word ``encryption'' is never explicitly presented to the user. This has both positive and negative consequences, as it minimizes interference and technical jargon, but may also prevent the user from realizing that the files are protected. Adding a ``More info'' link on the main page, leading to a brief description of the basic functions of \OmniShare, might be valuable for this reason. The addition of a security indicator icon could also remind the user that their files are safe.

In summary, our cognitive walkthrough evaluation showed that \OmniShare can be easily and safely used by novice users. We uncovered no fatal usability issues, and were able to fix the majority of usability issues identified in the cognitive walkthrough. We also identified two conceptual issues that affect not only \OmniShare, but also the design space of secure and password-less file sharing systems.


}

\section{Related Work}
\label{sec:relatedWork}

Solutions like SpiderOak~\cite{SpiderOak}, Wuala~\cite{Wuala} (now discontinued), and
Tresorit~\cite{Tresorit} offer secure cloud storage with client-side file
encryption. However they use keys derived from passwords to encrypt the files.
On the other hand, tools like Viivo~\cite{Viivo},
BoxCryptor~\cite{BoxCryptor}, and Sookasa~\cite{Sookasa} allow encryption with
client-generated keys and allow users to choose their preferred cloud storage. However, they use an
additional server to manage and distribute the file encryption keys across
devices. In contrast, the security of \OmniShare does not depend on any
server.

PanBox~\cite{panbox} is the closest solution comparable to \OmniShare. In
addition to client-side encryption, it uses OOB channels like Bluetooth and
Wi-Fi to distribute keys to client devices. 
However, this requires multiple user interactions, as described in the previous section. PanBox appears to be limited to German users. In
contrast \OmniShare delivers minimal, consistent user interaction and is
freely available to anyone.

\section{Conclusion}
\label{sec:conclusion}

Data privacy has become a major concern with respect to cloud storage. 
\OmniShare addresses this problem by combining client-side encryption with intuitive key distribution mechanisms.
The use of a key hierarchy and individual file keys facilitates selective synchronization of directories as well as sharing of files and directories with other users.
\OmniShare is open source software that is currently available for both Android and Windows, with other platforms under development.
As a generic mechanism to construct \emph{authorized device domains}, \OmniShare will have other applications beyond secure cloud storage. 
For example, suppose an online banking application uses trusted hardware on mobile devices to protect user credentials for online bank access. 
To allow the credentials to be used from multiple devices belonging to the same user, the application could allow the user to define an authorized domain of devices using \OmniShare and protect the banking credentials using the domain root key.
A similar approach could be used to synchronize encrypted passwords between the user's devices when using password managers such as LastPass.
Another promising avenue of future work is to consider how new technologies, such as fully homomorphic encryption or deduplication of encrypted data can be integrated into \OmniShare.

\section*{Acknowledgments}
This work was partially supported by the Academy of Finland project "Cloud Security Services (CloSe)" (Grant Number: 283135) and the Intel Collaborative Research Institute for Secure Computing (ICRI-SC).
\OmniShare has also received development funding as the overall winner of the \emph{Privacy via IT-Security} mobile app development competition at CeBIT 2016.
We thank Jan-Erik Ekberg, Brian McGillion, Jian Liu, and Alexandra Dmitrienko for their feedback on previous versions of this manuscript.



%

\ifCLASSOPTIONcaptionsoff
  \newpage
\fi



%
{
\raggedright
\bibliographystyle{IEEEtran}
\bibliography{references}
}

\ifanonymous

\else{

\fi




\ifsubmission
\else
\begin{appendices}
\clearpage
\newpage
\onecolumn

\section{Overview of Protocol Analysis using Scyther}
\label{sec:scyther}

Scyther~\cite{scyther} is an automated tool for reasoning about the security properties of message exchange protocols.
This section provides a brief introduction to the tool and its use in the analysis of the security protocols in \OmniShare.
Further details about this tool, as well as downloads, source code, and examples, are available from the tool's project page: \mbox{\url{https://www.cs.ox.ac.uk/people/cas.cremers/scyther/}}

Scyther is a \emph{symbolic} analysis tool in that it analyses symbolic representations of real protocols.
This type of analysis is well-suited for message exchange protocols since it focuses only on potential vulnerabilities arising from the protocol itself.
In a real deployment, other types of vulnerabilities, such as implementation bugs or vulnerabilities in the underlying platform, must also be considered, but are arguably orthogonal to vulnerabilities in the protocol.
The tool is \emph{automated} in that it only requires the user to provide an abstract specification of the protocol and the security properties of interest.
Based on this specification, the tool can determine which of the security properties hold, or provide counter-examples where properties do not hold.

The following appendices present the specifications for the two main protocols used in \OmniShare.
These are the complete specifications, and as such can be directly input to the Scyther tool (version 1.1.3) to reproduce our analysis results.

In Scyther specifications, a protocol consists of two or more \texttt{roles}, each representing a different type of communicating entity.
Multiple instances of each role may participate in any run of the protocol.
Each role has a number of local symbolic variables, which can be either freshly generated for each protocol run (denoted by the \texttt{fresh} keyword) or placeholders for information that will be received during the protocol (denoted by the \texttt{var} keyword).
The exchange of messages between roles is specified using the \texttt{send} and \texttt{recv} keywords, which have the following syntax:

\begin{lstlisting}[language=json,basicstyle=\footnotesize,frame=single]
send_[message_number] ( [sender], [recipient], [message (multiple comma-separated symbols)] );
recv_[message_number] ( [sender], [recipient], [message (multiple comma-separated symbols)] );
\end{lstlisting}

Scyther automatically models a Dolev-Yao style adversary, who has full control of the communication network (i.e. the adversary can eavesdrop, block, replay, modify, or forge any message).
The adversary may also take on one or more roles in the protocol.
The \texttt{[sender]} and \texttt{[recipient]} parameters in the send and receive operations thus only indicate the \emph{intended} recipient or the \emph{apparent} sender, since the adversary has complete control over which messages are received by which entities.

However, it is assumed that all cryptographic primitives are correctly implemented and cannot be subverted by the adversary.
Scyther can model symmetric and asymmetric cryptographic primitives, and includes a number of built-in keys that are assumed to have been pre-distributed among the relevant participants, in order to simplify the analysis.
Specifically, the keyword \texttt{k(A,B)} denotes a pre-shared key between roles A and B.
In the specifications of the \OmniShare protocols, we use this to model the out of band communication channel, which is assumed to be secure (i.e. confidential, integrity-protected, and authenticated).

After the message exchange has been specified, each role can include a number of \texttt{claim} statements, each of which captures a particular security property.
In the automated analysis, the tool attempts to prove or disprove each property.
The full details of these properties are described in~\cite{cremers2012operational}, but the following informal examples are given to provide an intuitive understanding of the \OmniShare protocol specifications that follow.

\begin{lstlisting}[language=json,basicstyle=\footnotesize,frame=single]
// Secrecy: the entity in role A claims that symbol KeySAuth is not known by the adversary
claim_a1(A, Secret, KeySAuth);

// Non-injective synchronization: the entity in role A claims that if has completed a run of the protocol, 
// the other entities with whom it believes it was communicating will agree that they have completed a 
// run of the protocol with this entity, and all entities will agree on the data items that were exchanged.
// This is used to model authentication.
claim_a2(A, Nisynch);

// Reachability: the entity in role A claims that there exists at least one sequence of events that will 
// allow it to reach this claim. This claim ensures that the message exchange protocol can be completed.
claim_a3(A, Reachable);
\end{lstlisting}

In addition to the claim statements, the \texttt{match} statement can be used to check whether a received symbol is equivalent to a local variable.
If the equivalence relationship is not satisfied, the protocol run terminates (which can subsequently be detected by the \texttt{Reachable} claim statement). 

\pagebreak

\section{Scyther Model of the Single Round-Trip Protocol}
\label{sec:scyther-single}

\begin{lstlisting}[language=json,firstnumber=1, basicstyle=\footnotesize, label={scyther:single-round-trip}]
hashfunction H;

/* 
 * Scyther assumes all agents have access to all built-in public keys. However,
 * our protocol does not have a pre-established Public key infrastructure (PKI).
 * Therefore we define an additional asymmetric key pair to model our key 
 * distribution protocol.
 */
const pk2: Function; // A function for public key which is different from the default Pk(A)
secret sk2: Function; // A corresponding secret key function
inversekeys (pk2,sk2); // Mapping of pk2 to sk2 allowing a value X encrypted with pk2 
                       // to be decrypted with sk2

protocol usingQR(A, B)
{ 
  role A {  
    fresh KeySAuth: Nonce; // a random session authentication key
    var RK: Nonce; // The root key
    var MAC: Ticket;
    fresh A1: Nonce; // nonce used to interpret asymmetric key in Scyther.
    
    // Messge sent via secure out of band communication channel 
    // (modelled using built-in key shared between A and B)
    send_0(A, B, {KeySAuth, H(pk2(A1))}k(A, B));     

    // Messages sent via cloud communication channel
    send_1(A, B, pk2(A1));       
    macro m = {RK}pk(A);
    recv_2(B, A, (m, MAC)); 
    
    // verification of the integrity of encrypted root key
    match(MAC, H(m, KeySAuth));

    // Claims
    claim_a1(A, Secret, sk2(A1));
    claim_a2(A, Secret, KeySAuth);
    claim_a3(A, Secret, RK);
    claim_a4(A, Nisynch);
    claim_a5(A, Reachable);
  } 

  role B {    
    var hash: Ticket;
    var KeySAuth: Nonce;
    fresh RK: Nonce;
    var A1: Nonce; 

    // Messge received via secure out of band communication channel 
    // (modelled using built-in key shared between A and B)
    recv_0(A, B, {KeySAuth, hash}k(A, B));

    // Message received via cloud communication channel
    recv_1(A, B, pk2(A1));
    // OOB message verfication
    match(hash, H(pk2(A1)));  
    
    //Sending back the encrypted rootkey and an HMAC for integrity protection.    
    macro m = {RK}pk(A);
    send_2(B, A, (m, H(m, KeySAuth)));
    
    // Claims
    claim_b1(B, Secret, KeySAuth);
    claim_b2(B, Secret, RK);  
    claim_b3(B, Nisynch);
    claim_b4(B, Reachable);
  }
}
\end{lstlisting}

\pagebreak

\section{Scyther Model of the Multiple Round-Trip Protocol}
\label{sec:scyther-multiple}

\begin{lstlisting}[language=json,firstnumber=1, basicstyle=\footnotesize, label={scyther:multiple-round-trip}]
hashfunction g1, g2, H;
function f, plus;

/* 
 * Support protocol for simulating modular exponential equivalent
 */
protocol @exponentiation(BE, BM1, AM2)
{
  role BE {
    //Simulation of (g^a)^b = (g^b)^a
    var a, b: Ticket;

    recv_!1(BE, BE, g2(g1(a), b));
    send_!2(BE, BE, g2(g1(b), a));
  }
  
  /*
   * Support protocol to simulate equality of M1 and M1'
   */
  role BM1 {
    var alpha, beta, a, b, x: Ticket;
    recv_!3(BM1, BM1, H(alpha, beta, H(f(g2(g1(b), a), g2(g1(b), x)))));
    send_!4(BM1, BM1, H(alpha, beta, H(f(g2(g1(a), b), g2(g1(x), b)))));
  }

  /*
   * Support protocol to simulate equality of M2 and M2'
   */
  role AM2 {
    var alpha, beta, a, b, x, RK: Ticket;
    macro KeySesA = f(g2(g1(b), a), g2(g1(b), x));
    macro KeySesB = f(g2(g1(a), b), g2(g1(x), b));
    macro M1A = H(alpha, beta, H(KeySesA));
    macro M1B = H(alpha, beta, H(KeySesB));
    macro M2A = H(alpha, M1A, KeySesA);
    macro M2B = H(alpha, M1B, KeySesB);
    recv_!5(AM2, AM2, {RK}KeySesB, M2B);
    send_!6(AM2, AM2, {RK}KeySesA, M2A);
  }
}

/*
 * Authorization protocol using passcode based on SRP
 */
protocol usingPasscode(A, B)
{
  role A {
    fresh P, a: Nonce;
    var s, gb, v, RK: Ticket;
    macro x = H(s,P);
    macro alpha = g1(a);
    macro beta = plus(gb, v);
    macro KeySes = f(g2(gb, a), g2(gb, x));
    macro M1 = H(alpha, beta, H(KeySes));
    macro M2 = H(alpha, M1, KeySes);
    
    // Message sent via the secure out of band communication channel
    // (modelled by encrypting with the default shared key between A and B).
    send_0(A, B, {P}k(A,B));    
    
    // Messages exchanged via the cloud channel
    send_1(A, B, alpha);
    recv_2(B, A, beta, s);
    
    match(v, g1(x));
    send_!3(A, B, M1); // Sending M1
    recv_!4(B, A, {RK}KeySes, M2); // Receiving M2, from support protocol instead of role B

    claim_a1(A, Reachable);
    claim_a2(A, Secret, a);
    claim_a3(A, Secret, P);
    claim_a4(A, Secret, KeySes);
    claim_a5(A, Secret, RK);
    claim_a6(A, Nisynch);
  }

  role B {
    fresh b, s, RK: Nonce;
    var alpha, s, P;
    macro x = H(s,P);
    macro beta = plus(g1(b), g1(x));
    macro KeySes = f(g2(alpha, b), g2(g1(x), b));
    macro M1 = H(alpha, beta, H(KeySes));
    macro M2 = H(alpha, M1, KeySes);
    
    recv_0(A, B, {P}k(A, B));    
    
    recv_1(A, B, alpha);
    send_2(B, A, beta, s);
    recv_!3(A, B, M1); // Receiving M1, from the support function instead of role A 
    send_!4(B, A, {RK}KeySes, M2); //Sending out M2
    
    claim_b1(B, Reachable);
    claim_b2(B, Secret, b);
    claim_b3(B, Secret, P);
    claim_b4(B, Secret, KeySes);
    claim_b5(B, Secret, RK);
    claim_b6(B, Nisynch);
  }
}
\end{lstlisting}

\end{appendices}
\fi

\end{document}